\documentclass[aps,prl,preprint,amsbsy,amssymb]{revtex4}
\usepackage{indentfirst,latexsym,bm,amsmath,graphicx}

\begin{document}

\title{On the (circular) polarization-independence of microwave-induced resistance oscillations
       and zero resistance state}
\author{Shenshen Wang and Tai-Kai Ng}

\address{
 Department of Physics, Hong Kong University of Science and
 Technology, Clear Water Bay Road, Kowloon, Hong Kong}

\date{ \today }

\begin{abstract}

  The immunity of microwave-induced magneto-resistance oscillations and corresponding zero resistance regions to
  the direction of (circular) polarization of microwave is studied in this paper. We propose that a spontaneous
  circular motion of the whole electron fluid would stabilize the system and minimize the polarization
  sensitivity of the oscillatory DC resistance. Results of a self-consistent calculation capture the qualitative
  features of the experimental observation.

\end{abstract}


\maketitle

 The observation of ``Zero-(DC)-Resistance State" (ZRS) in high mobility two-dimensional electron systems (2DES)
 under magnetic field and microwave (MW) radiation presented a surprise to the physics community
 \cite{Mani02,Zudov03}. When the system, under crossed uniform magnetic field and small dc bias, is irradiated by
 microwave of sufficient intensity, the longitudinal (dissipative) resistance $R_{xx}$ develops strong
 oscillatory dependence on the magnetic field. At low temperature and high radiation power the
 minima of the oscillations evolve into ZRS.

 Intense interests in the physics community have been aroused by this unusual non-equilibrium phenomenon.
 Theoretical efforts aimed at identifying the microscopic mechanisms accounting for the MW-induced resistance
 oscillations (MIRO) and ZRS. Early theories employed the picture of photon-excited transport assisted by
 short-range scattering\cite{Durst03,Vavilov&Aleiner,CM}, and later the scenario of MW-induced oscillations in
 the nonequilibrium electron distribution function was argued to be the leading cause of MIRO\cite{distribution
 function}. Other mechanisms based on different physical pictures were also proposed\cite{tunnelling,gap,
 plasma,non-parabolicity}. As for
 ZRS, though there exist proposals that do not invoke negative resistance (NR) instability \cite{coherent}, the
 pattern formation model \cite{NRI}, which is based on NR, remains the most popular. On the experimental side,
 the outstanding issues are activated temperature dependence with large energy gaps \cite{T-dep}, immunity of
 MIRO and ZRS to the polarity of circular polarization of MW \cite{PI}, suppression of MIRO and ZRS by
 in-plane magnetic fields \cite{in-plane B field}, and multi-photon processes in the ZRS formation
 \cite{muitiphoton}; most of these findings are not readily accommodated by existing theories.

 Among other issues, the polarization immunity of DC resistance poses a particularly challenging
 test. Experiment by Smet {\em et al.}\cite{PI} established that MIRO are {\em insensitive} to the polarization
 state of the MW radiation. However, transmission data also shows that active cyclotron resonance absorption occurs
 {\em only} when the circular polarization matches the magnetic field orientation (denoted as CRA), although ZRS
 is observed under both MW polarities. This observation was inconsistent with the two most prevailing
 theories, the impurity and/or phonon assisted inter-Landau level (LL) transitions model\cite{Durst03,
 Vavilov&Aleiner,CM} and the non-equilibrium distribution function scenario\cite{distribution function}. These
 theories predict oscillations with correct period and phase, however, with {\em substantially
 different} amplitudes for the different polarizations; the MW photoconductivity of the cyclotron resonance
 inactive (CRI) state is smaller by a factor $\left((\omega-\omega_c)^2+\Gamma^2\right)/\left((\omega+\omega_c)^2
 +\Gamma^2\right)$, where $\Gamma^{-1}$ is a phenomenological lifetime. The factor reflects the huge difference in
 the (AC) Drude conductivity for opposite circular polarities.

 Polarization immunity of the DC resistance posts a big challenge to the understanding of the ZRS. To
 produce the ``same" MIRO and ZRS, there must be an additional mechanism that compensates for the discrepancy in the
 energy absorption rate between the CRA and CRI states. In this paper, we propose that a spontaneous circular
 motion of the whole electron fluid previously proposed by Ng and Dai \cite{TK05} could stabilize the system and
 minimize the polarization dependence of the oscillatory photoconductivity. The spontaneous circular motion
 arises in the cyclotron resonance (CR) favorable orientation whenever the intensity of incident MW radiation
 exceeds a threshold value.

 To see how this could happen, we examine the general transport equation for the center of mass (CM) coordinate $\vec{R}(t)$ of
 the electron liquid where the effect of impurity is included to the second order \cite{TK05}:
\begin{equation}
\label{e-motion}
 m\ddot{\vec{R}}(t)=(-e)\left(\vec{E}(t)+\frac{1}{c}\dot{\vec{R}}(t)\times\vec{B}\right)
 +\alpha\nabla_{\vec{R}(t)}\int\limits_{-\infty}^t\mathrm{d}
 t'\chi\left(\vec{R}(t)-\vec{R}(t');t-t' \right),
\end{equation}
 where $\alpha=n_i|u|^2/\bar{n}$, $n_i$ is the density of impurity, $|u|^2$ indicates the strength of the impurity potential,
 $\bar{n}=N/V$ is the carrier density, and $\chi$ is the (equilibrium) retard density-density response function
 without the MW term. The equation is derived in the CM frame where the MW field is eliminated and the
 electron liquid sees moving impurities following the path $\sim-\vec{R}(t)$. After impurity averaging, we obtain
 Eq.\ (\ref{e-motion})\cite{TK05}. We note that similar equation has also been
 proposed by Lei \textit{et al.}\cite{CM}. The electric field relevant to this phenomenon is of form
 $\vec{E}(t)=\vec{E}_1\cos\left(\bar{\omega}t\right)+\vec{E}_2\sin\left(\bar{\omega}t\right)+\vec{E}_d,$
 where $\vec{E}_2=\pm\hat{z}\times\vec{E}_1$ and $\left|\vec{E}_1\right|=\left|\vec{E}_2\right|=E_0.$
 The first two terms represent a circularly-polarized MW radiation with frequency $\bar{\omega}$ and $\vec{E}_d$
 stands for a small DC bias. The plus/minus sign in the expression of $\vec{E}_2$ indicates CRA/CRI state. For
 small DC bias $\vec{R}(t)$ can be written as $\vec{R}(t)=\vec{R}_{AC}(t)+\vec{R}_{DC}(t),$ where
 $\vec{R}_{AC}(t)$ is the (dominant) part induced by MW field and $\vec{R}_{DC}(t)\sim\vec{v}t$ is a small
 DC correction.

 Eq.\ (\ref{e-motion}) suggests that the linear DC resistance is determined completely by $\vec{R}_{AC}(t)$
 since there is nothing else in the equation. As sketched in Fig.1 (left solid circles), $\vec{R}_{AC}(t)$ for
 opposite polarization states have very different amplitudes of motion in the Drude model, and to explain the
 polarization immunity, this big difference in $\vec{R}_{AC}(t)$ must be minimized. A plausible way for this to
 occur is to have a spontaneous circular motion in the CRA direction for both CRA/CRI states. If this
 spontaneous circular motion dominates over the Drude motion, the difference in $\vec{R}_{AC}(t)$ between the
 CRA and CRI states will be minimized. In reality the composite trajectory of the spontaneous and Drude motion can be
 rather complicated (Fig.1 right-hand solid traces) and a self-consistent numerical calculation has to be
 performed to test this idea.

 \begin{figure}[htb]
 \centerline{\includegraphics[scale=0.5]{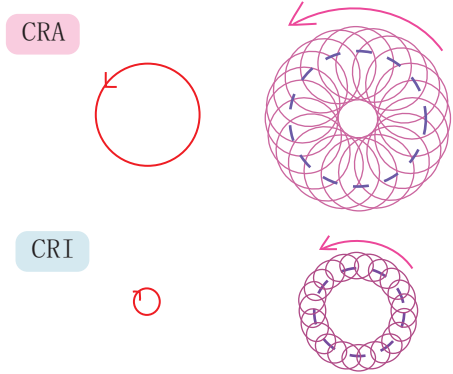}}\caption{Schematic trajectory of the AC motion of
  the electron fluid in the absence (left) and presence (right) of the spontaneous circular motion (dashed circle)
  for CRA (upper) and CRI (lower) states.}
 \end{figure}

 To test our proposal, we study a trial solution of Eq.\ (\ref{e-motion}) with
\begin{equation}
\vec{R}(t)=\vec{v}\cdot t+\vec{R}_f(t)+\vec{R}_s(t)=\vec{v}\cdot
t+\vec{A}_1\cos\left(\bar{\omega}t+\delta\right)+\vec{B}_1\sin\left(\bar{\omega}t+\delta\right)
+\vec{\alpha}_1\cos\left(\omega_{_I}t\right)+\vec{\beta}_1\sin\left(\omega_{_I}t\right),
\end{equation}
 where $\vec{R}_f(t)$ describes the CM motion directly coupled to the radiation field and $\vec{R}_s(t)$ depicts
 the spontaneously generated oscillatory mode with frequency $\omega_{_I}$. We shall call them ``fast" and
 ``slow" modes in the following in view of their frequency difference which will be shown numerically later.
 $\delta$ is the phase delay of the ``fast mode" with respect to the AC driving. Notice that we keep
 only the base harmonic modes in $\vec{R}_f(t)$ and $\vec{R}_s(t)$. This can be justified for the ``fast" mode
 since in the weak radiation limit the size of the circular orbit $R_c\sim\sqrt{\vec{A}_1^2+\vec{B}_1^2}$ will be
 much less than the magnetic length $l_c=\sqrt{c/eB}$ (We set $\hbar=1$ in the following) which is quite long for
 the B field ($\sim1 \mathrm{T}$) relevant to this phenomenon\cite{TK05}. Notice that such an argument does not apply to
 the ``slow" mode whose amplitude is not governed by the radiation strength.

 We next expand the impurity-scattering induced damping force
 $\vec{F}_{_I}(t)\equiv\alpha\nabla_{\vec{R}(t)}\int\limits_{-\infty}^t\mathrm{d}t'\chi
 \left(\vec{R}(t)-\vec{R}(t');t-t'\right)$
  in a Bessel-Fourier series and keep only the base-frequency oscillating terms. By putting the
 approximate $\vec{R}(t)$ and $\vec{F}_I(t)$ into Eq. (1) and compare, we obtain three force-balance equations
\begin{eqnarray}
0&=&-e\vec{E}_d-\frac{e}{c}\;\vec{v}\times\vec{B}+\vec{F}_v,\\
m\ddot{\vec{R}}_f(t)&=&-e\vec{E}(t)-\frac{e}{c}\;\dot{\vec{R}}_f(t)\times\vec{B}+\vec{F}_{_I}^{(f)}(t),\\
m\ddot{\vec{R}}_s(t)&=&-\frac{e}{c}\;\dot{\vec{R}}_s(t)\times\vec{B}+\vec{F}_{_I}^{(s)}(t).
\end{eqnarray}
Here
\begin{equation}
\vec{F}_v=-\alpha\iint\mathrm{d}^dq\:\vec{q}\sum_{m,n=-\infty}^{\infty}Im\chi
\left(\vec{q},\vec{q}\cdot\vec{v}+m\bar{\omega}+n\omega_{_I}\right)J_m^2\left(z\left(\vec{q}\right)\right)
J_n^2\left(y\left(\vec{q}\right)\right)
\end{equation}
 is the time-averaged damping force, where $z(\vec{q})^2=(\vec{q}\cdot\vec{A}_1)^2+(\vec{q}\cdot\vec{B}_1)^2$
 and $y(\vec{q})^2=(\vec{q}\cdot\vec{\alpha}_1)^2+(\vec{q}\cdot\vec{ \beta}_1)^2$. For the ``fast mode",
 $\vec{F}_{_I}^{(f)}(t)\sim\vec{H}_1\cos\left(\bar{\omega}t+\delta\right)+\vec{G}_1
 \sin\left(\bar{\omega}t+\delta\right)$ where $\vec{H}_1\equiv\pi^{(1)}\cdot\vec{A}_1-\pi^{(2)}\cdot\vec{B}_1$
 and $\vec{G}_1\equiv\pi^{(1)}\cdot\vec{B}_1+\pi^{(2)}\cdot\vec{A}_1$ whereas for the ``slow mode",
 $\vec{F}_{_I}^{(s)}(t)\sim\vec{h}_1\cos\left(\omega_{_I}t\right)+\vec{g}_1\sin\left(\omega_{_I}t\right)$,
 where $\vec{h}_1\equiv\pi^{(1)'}\cdot\vec{\alpha}_1-\pi^{(2)'}\cdot\vec{\beta}_1$
 and $\vec{g}_1\equiv\pi^{(1)'}\cdot\vec{\beta}_1+\pi^{(2)'}\cdot\vec{\alpha}_1$.

 The $\pi's$ are given by
 \begin{eqnarray}
 \pi^{(1)}&=&\frac{\alpha}{2}\iint\mathrm{d}^dq\;q^2\sum_{m,n=-\infty}^{\infty}
 Re\chi\left(\vec{q},\;\vec{q}\cdot\vec{v}+m\bar{\omega}+n\omega_{_I}\right)
 \left[J_m(z(\vec{q}))\frac{J'_m(z(\vec{q}))}{z(\vec{q})}\right]J_n^2(y(\vec{q})),\nonumber\\
 \pi^{(2)}&=&\frac{\alpha}{2}\iint\mathrm{d}^dq\;q^2\sum_{m,n=-\infty}^{\infty}
 Im\chi\left(\vec{q},\;\vec{q}\cdot\vec{v}+m\bar{\omega}+n\omega_{_I}\right)
 \;m\!\left[\frac{J_m(z(\vec{q}))}{z(\vec{q})}\right]^2J_n^2(y(\vec{q})),
 \end{eqnarray}
 and $\pi^{(1)'}$, $\pi^{(2)'}$ can be obtained from $\pi^{(1)}$, $\pi^{(2)}$ by simply interchanging $z\leftrightarrow y$,
 $m\leftrightarrow n$ and $\bar{\omega}\leftrightarrow\omega_{_I}$. We shall consider the limit $\vec {v}\rightarrow 0$
 when calculating $\pi's$ in the following since we are interested in the linear-response DC current.

 After some algebra we obtain the equation of motion for "fast mode"
 \begin{equation}
 \sqrt{\left[m\bar{\omega}^2\left(1\mp\omega_c/\bar{\omega}\right)+\pi^{(1)}\right]^2+\left[\pi^{(2)}\right]^2}\cdot A_1=eE_0,
 \end{equation}
 where $-/+$ corresponds to CRA/CRI state. The equation is of Drude form representing an oscillation driven
 by AC electric field under impurity-induced friction $\pi^{(2)}$ that depends on the amplitudes of both ``fast"
 and ``slow" modes $\left(z(\vec{q})\ \mbox{and}\  y(\vec{q})\right)$. A corresponding reactive correction $\pi^{(1)}$
 also appears which can be interpreted as a mass correction\cite{TK05}. The phase delay $\delta$ is given by
 $\tan\delta=\pi^{(2)}/\left[m\bar{\omega}^2(1\mp\omega_c/\bar{\omega})+\pi^{(1)}\right]$. Notice the explicit
 polarization dependence of $A_1$ in Eq. (8).

 Since there is no external driving force, the spontaneously-generated ``slow" mode is determined by the
 self-sustainability requirement that the corresponding frictional force vanishes, i.e.
\begin{equation}
\pi^{(2)'}=0.
\end{equation}
 Putting it into Eq.(5) we obtain another equation for ``slow mode",
 \begin{equation}
 -m\omega_{_I}^2\left(1-\omega_c/\omega_{_I}\right)\alpha_1=\pi^{(1)'}\!\cdot\alpha_1.
 \end{equation}
 The two equations determine self-consistently the amplitude and frequency of the ``slow mode". Notice that the
 spontaneous circular motion is always in the CR-favorable direction, independent of the
 polarization of the MW radiation. Polarization dependence enters only indirectly through $z(\vec{q})$
 which appears in both $\pi^{(1)'}$ and $\pi^{(2)'}$.

 Eq.s (8)--(10) form a set of self-consistent equations determining the ``fast" and ``slow" mode amplitudes $A_1$,
 $\alpha_1$ and the ``slow" mode frequency $\omega_{_I}$. These equations are solved numerically. The
 longitudinal DC resistance $R_{xx}=-\lim\limits_{\vec{v}\rightarrow0}\left[ v^{-2}\vec{F}_v\cdot\vec{v}\right]$
 is computed afterward.

 We have employed the density-density response function of non-interacting electron gas in constant magnetic
 field\cite{response function} in our calculation, with Landau levels (LLs) broadened
 phenomenologically into Lorentzians, i.e.
 $\delta(\varepsilon-n\omega_c)\rightarrow \pi^{-1}\Gamma/\left((\varepsilon-n\omega_c)^2+\Gamma^2\right)$.
 Since experimentally $\bar{\omega}\sim 200\mathrm{GHz}$ is fixed while sweeping the B field, we use $\bar{\omega}$ as the basic
 unit and set $T\sim\bar{\omega}$, $\Gamma(T)\sim0.2\bar{\omega}$,
 $E_{_F}\sim10\bar{\omega}$ and keep 20 LLs in our calculation, consistent with the low field
 ($B\sim1\mathrm{T}$) and intermediate temperature ($T\sim1\mathrm{K}$) setting in experiment. We shall vary magnetic field $B$
 and use the frequency ratio $\omega_n\equiv\bar{\omega}/\omega_c$ as abscissa in presenting our results. We also
 define normalized radiation intensity $I_{_N}\equiv\left(eE_0/m^{*}\bar{\omega}^2\bar{l}\,\right)^2$ where
 $\bar{l}\equiv\sqrt{\omega_c/\bar{\omega}}\:l_c$ ($m^*=0.068m_0$ is the effective mass of conduction band electrons in GaAs)
 as well as normalized amplitudes $c\equiv A_1/\,\bar{l}$, $d\equiv\alpha_1/\,\bar{l}$
 and renormalized frequency $x\equiv\omega_{_I}/\omega_c$ in our calculation.

 To further simplify calculation and analysis, we make another approximation of keeping only $|n|,|m|\leq1$
 terms (zero- and one-photon processes) in the impurity induced forces in our calculation; this is consistent
 with keeping only base harmonics in our trial trajectory. Correspondingly, we restrict our calculation to the
 frequency range $\omega_n=\bar{\omega}/\omega_c\sim1$--$2$ since MIRO occur at the weak B field side
 $\bar{\omega}>\omega_c$ and higher frequency range will be dominated by transitions between higher LLs
 ($|n|,|m|>1$ processes). The approximation of keeping only single photon process in the "fast mode" can be
 justified in the weak radiation field limit. However the approximation can not be justified {\em a prior} for the
 spontaneous (slow) mode. We have estimated the effect of multi-photon processes associated with the ``slow mode"
 on our self-consistent equations and found that it does not affect terms associated with
 $Im\chi(\vec{q},\:\vec{q}\cdot\vec{v}+n\omega_{_I}+\bar{\omega})$ but
 would enhance terms associated with $Im\chi(\vec{q},\:\vec{q}\cdot\vec{v}+n\omega_{_I})$ and
 $Im\chi(\vec{q},\:\vec{q}\cdot\vec{v}+n\omega_{_I}-\bar{\omega})$. To mimic these effects we introduce a
 correction factor $a>1$ that multiplies the latter two terms in $\pi^{(2)'}$. The same factor is introduced
 to the counterparts in $\pi^{(1)'}$ and $\vec{F}_v$ ($R_{xx}$) to ensure consistency in our calculation.
 As long as $a$ does not deviate from $a=1$ too much, the qualitative behavior of the solutions is not sensitive
 to $a$. We shall present our calculation results with $a=1.2$ in the following.

 Numerically we find that we may roughly divide the region $\omega_n\sim1$--$2$ into negative resistance (NR) region
 that centers about $\omega_n\sim1.2$, and positive resistance (PR) region that peaks at $\omega_n\sim1.7$.
 For radiation intensity $(I_{_N})_{th}<I_{_N}<(I_{_N})_{multi-\bar{\omega}}$, where $(I_{_N})_{th}\!\sim\!0.001$
 numerically (corresponding to $E_0\sim7\,\mathrm{V/cm}$) is a threshold value above which ``slow" mode
 appears and $(I_{_N})_{multi-\bar{\omega}}\sim0.02$ ($E_0\sim32\,\mathrm{V/cm}$) is a value where multi-photon
 processes of MW radiation become important, we find that the solutions for $c$, $d$ and $x$ fall within the
 ranges listed in Table 1.
 \begin{center}
 \begin{table}[htb]
 \begin{tabular}{|c|c|c|c|c|}\hline
 \;\textbf{polarity}&    \;\textbf{frequency range}\,&   \, $\,\boldsymbol{d}$\,&  $\boldsymbol{c}$&     $\boldsymbol{x}$\,\\ \hline
  $<${CRA}$>$&             NR($\omega_n:1.1$--$1.5$)&   \, $0.1$--\,$0.2$\,&        $0.2$--\,$0.4$&     $\;\sim0.3$\,\\ \hline
             &             PR($\omega_n:1.5$--$1.9$)&   \, $0.2$--\,$0.3$\,&        $0.2$--\,$0.4$&     $\;\sim1.1$\,\\ \hline
  $<${CRI}$>$&                ($\omega_n:1.1$--$2.0$)&   \, $\sim0.1$\,&          \,$0.03$--\,$0.08$\,&  $\;\sim0.3$\,\\
  \hline
 \end{tabular}
 \caption{The numerical range of the self-consistently determined $d$, $c$ and $x$ for CRA and CRI
 states in the regime $\omega_n\sim1$--$2$.}
 \end{table}
 \end{center}
 Notice that the amplitude of ``fast mode" ($c$) is much smaller in the CRI state when compared with CRA state as
 expected. However, the amplitudes $d$ of the ``slow mode" in the two polarization states are comparable. The
 low-frequency ``slow mode" ($\omega_{_I}\!\sim\!0.3\omega_c$) appears in the whole frequency range of CRI state and
 the NR region of CRA state. Unexpectedly, $d\neq0$ solutions also exist at the PR region of CRA state with
 rather large amplitude and near-CR frequency.

 The effect of the ``slow mode" on the polarization-dependence of DC resistance and ZRS can be roughly understood
 as follows: when the incident radiation is sufficiently strong and NR instability shows up, a spontaneous
 circular motion is generated in the electron fluid in CR favorable direction. Mathematically, low-frequency
 fictitious photons represented by $n\neq0$ processes are spontaneously generated and additional photon-assisted
 transport channels open up. As a result normal dissipation is enhanced and NR is cured. This mechanism
 dominates the whole frequency range of CRI state and the NR region of CRA state. The spontaneous circular
 motion also turns the
 original fast-rotating CRA orbit (left-hand solid circle of Fig.1) into the ``lace" of a slowly-rotating
 ($\omega_{_I}$) orbit (right-hand dashed circle). The strong mixing of the two orbits destroys the
 identity of well-defined Landau orbit and suppresses photon-assisted scattering in the CRA state, thus
 drawing the two states more close to each other. Mathematically, the suppression comes through
 the increase in the argument ($y(\vec{q})$) of Bessel functions.

 \begin{figure}[htb]
 \centerline{\includegraphics[scale=0.7]{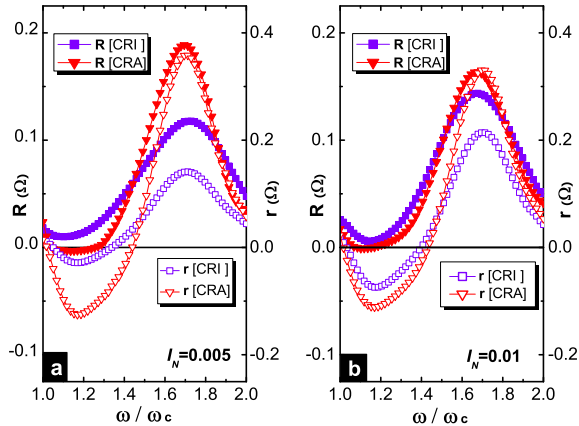}}\caption{DC resistance versus frequency ratio for CRA and CRI states
  under two values of radiation intensity: (a) $I_{_N}=0.005$ (moderate intensity) and (b) $I_{_N}=0.01$ (high
  intensity). ($R$)--- with slow mode included in calculation, ($r$)--- slow mode is excluded.$\quad$}
 \end{figure}

 We show in Fig. 2 the calculated DC resistance versus frequency ratio both with ($R$) and without ($r$) ``slow
 mode" included in our model. Notice the different plotting scale for the two calculated resistances in the figure. We
 see that when the radiation is of moderate intensity $I_{_N}\!=0.005$($E_0\sim16\,\mathrm{V/cm}$)
 (Fig.2a) , NR region becomes pronounced in the absence of the ``slow mode" for both CRA and CRI states and the
 disparity in the oscillatory amplitude of the two polarization states is large.

 The entering of the ``slow mode" suppresses the oscillatory amplitudes, especially for CRA state, which
 effectively reduces the amplitude discrepancy between the two polarizations; meanwhile the NR instability
 is almost completely healed by the $\omega_{_I}$-photon-excited inter-LL transitions in our highly simplified
 model. As the radiation strength further increases to $I_{_N}\!=0.01$($E_0\sim22\,\mathrm{V/cm}$) (Fig.2b),
 oscillation becomes stronger for CRI state but is approaching saturation for CRA state in the absence of the
 ``slow mode". In this case, the suppression effect plus the photon-assisted transport processes associated with
 the ``slow mode" make the oscillatory shape of the two polarization states even closer and lift the NR region to
 ZR for both polarities. We note also that at the B field region near CR where significant absorption takes place,
 no $d\neq0$ solution is found in our calculation and the oscillation amplitude of CRA and CRI
 curves are considerably different, in agreement with experiment \cite{PI}.

 In conclusion, we propose and demonstrate numerically that a spontaneous slowly-rotating circulating current in
 a 2DES under magnetic field will be generated when the incident circularly-polarized MW radiation exceeds certain threshold
 intensity. The spontaneous mode can cure the NR problem and provides a plausible explanation for the observed
 ZRS and polarization immunity of the DC resistance. Our calculation is crude because of the many approximations
 we made and the results are only in semi-quantitative agreement with experiment. Nevertheless, we believe our
 theory has provided a promising starting point to understand the physics behind the ZRS phenomenon
 and associated polarization immunity.

   We acknowledge support from HKUGC through grant CA05/06.SC04.

\end{document}